# Polytypism in LaOBiS$_2$-type compounds based on different three-dimensional stacking sequences of two-dimensional BiS$_2$ layers

Qihang Liu, Xiuwen Zhang, Alex Zunger

University of Colorado, Boulder, Colorado 80309, USA

**Abstract**

LaOBiS$_2$-type materials have drawn much attention recently because of various interesting physical properties, such as low-temperature superconductivity, hidden spin polarization, and electrically tunable Dirac cones. However, it was generally assumed that each LaOBiS$_2$-type compound has a unique and specific crystallographic structure (with a space group *P*4/*nmm*) separated from other phases. Using first-principles total energy and stability calculations we confirm that the previous assignment of the centrosymetric *P*4/*nmm* structure to LaOBiS$_2$ is incorrect as a phonon instability renders this structure impossible. Furthermore, we find that the unstable structure is replaced by a *family* of energetically closely spaced modifications (polytypes) differing by the layer sequences and orientations. We find that the local Bi-S distortion leads to three polytypes of LaOBiS$_2$ with different stacking patterns of the distorted BiS$_2$ layers. The energy difference between the polytypes of LaOBiS2 is merely ~1 meV/u.c., indicating the possible coexistence of all polytypes in the real sample and that the particular distribution of polytypes may be growth-induced. The in-plane distortion can be suppressed by pressure, leading to a phase transition from polytypes to the high-symmetry *P*4/*nmm* structure with a pressure larger than 2.5 GPa. In addition, different choices of the intermediate atoms (replacing La) or active atoms (BiS$_2$) could also manifest different ground state structures. One can thus tune the distortion and the ground state by pressure or substituting covalence atoms in LaOBiS$_2$-family.

*I. Introduction*

LaOBiS$_2$ represents a group of materials consisting of BiS$_2$ layers whose states dominate the low-energy spectrum, separated by electronically passive La$_2$O$_2$ buffer layers. LaOBiS$_2$-type materials have drawn attention because of the measured superconductivity[1-14] (up to $T_c$ = 10.6 K) and because the BiS$_2$ layer is spin-orbit active and thus leads to many interesting spin-related properties, including hidden spin polarization[15], spin field effect transistor[16] and electrically tunable Dirac cones[17]. The general crystal structure consists of a stack of layers (BiS$_2$)$^-$/(La$_2$O$_2$)$^{2+}$/(BiS$_2$)$^-$ made of two BiS$_2$ rocksalt-like layers and an intermediate fluorite-like La$_2$O$_2$ buffer layer separating them[18]. Many materials described by this structural scheme have recently been synthesized[1, 2, 4, 5, 7, 9, 10, 14], and even more members not listed in inorganic compound data bases (ICSD) can be conceived[17, 19]. As with many other compounds, here too it was generally assumed that each compound has a unique and specific crystallographic structure separated from other phases. Using first-principles calculations we find that contrary to this view[1, 6, 8, 11, 12, 18] the three-dimensional (3D) structure of this important family of compounds represents instead a family of energetically closely spaced modifications differing by the layer sequences and orientations. Such *polytypes* are "natural superlattices", familiar from other areas in condensed matter such as SiC[20, 21], ZnX (X = S, Se, Te)[22-24], and ATiO$_3$ (A = Sr and Ba) perovskites[25, 26]. Polytypes hold a special place as structural phases in that they have very similar (within meV) total energies, yet are distinguished by those physical properties that are sensitive to orientation and stacking such as polarization of light emission and sometimes the magnitude of the zone-folded band gap (for example, the optical band gap of cubic vs wurtzite polytype of SiC differ by nearly 1 eV[29]). What makes LaOBiS$_2$ polytypes discovered here particularly interesting is that (i) the characteristic local structural distortions in the electronically active BiS$_2$ layers could lead to charge fluctuations which may related to the electron-phonon coupling and the mechanism of superconductivity[1, 30]; and (ii) different polytypes are either centrosymmetric or noncentrosymmetric, a distinction that carries with it different physical properties related to presence or absence of inversion symmetry, such as band splitting, second harmonic generation and (hidden) spin polarization[15].

In this article we use first-principles total energy calculations to assess the structure and stability of various polytypes in the LaOBiS$_2$-family of compounds. We find that the local Bi-S distortion leads to three polytypes of LaOBiS$_2$ according to the different stacking pattern of distorted BiS$_2$ layers, rather than the previously assumed *P4/nmm* structure. The in-plane distortion is mainly induced by the interaction between the intermediate LaO and the BiS$_2$ layers, and the electron hybridization in the BiS$_2$ layer. We performed structural optimization in steps starting from the most restricted degrees of freedom to show how a small distortion can transform the assumed unstable structure to its three polytypes. Since the energy difference between the polytypes of LaOBiS2 is merely ~1 meV/u.c., the real 3D material constructed by stacking up such three types of building blocks could manifest various sequences of polytypes depending sensitively on the growth condition. Furthermore, the in-plane distortion can be suppressed by pressure, leading to a phase transition from polytypes to the high-symmetry *P4/nmm* structure with a pressure larger than 2.5 GPa. In addition, different choices of the buffer atoms (replacing La) or active atoms (BiS$_2$) manifest different ground state polytypes.

## *II. The in-plane structure of the 2D BiS$_2$-like layer and the emergence of possible stacking sequences*

LaOBiS$_2$[18] has been generally accepted to have the ZrCuSiAs-type structure, with a centrosymmetric *P4/nmm* space group. X ray diffraction data was refined[1, 6] by fitting the measured intensities to this structure type (named T$_0$ hereafter). In T$_0$ the BiS$_2$ layers in the (BiS$_2$)$^-$/(La$_2$O$_2$)$^{2+}$/(BiS$_2$)$^-$ stacks have four equal Bi-S bonds, as shown in Fig. 1. However, recent phonon calculations using density functional theory (DFT) found that the T$_0$ structure with *assumed* equal Bi-S bond lengths of 2.89 Å has soft phonons at the Brillouin zone center Γ point, which indicates dynamic instability and the theoretical impossibility of this T$_0$ structure[11] being stable. Yildirim showed that by shifting the S atoms along the diagonal of the 2D square (*x* or *y* direction in Fig. 1b), forming a ferroelectric-like distortion with two different in-plane Bi-S bonds (bond lengths 2.80 Å and 2.98 Å, see Fig. 1c) one can remove the dynamic instability. In this calculation a deformation parameter Q, which is defined by the normal modes having the unstable phonon was introduced to represent the in-plane displacement of S atoms. The total energy was minimized with respect to Q, with the remaining structural degrees of freedom other than Q (such as lattice parameter and atomic positions) kept frozen at their Q = 0 value (corresponding to the $T_0$ structure). A subsequent neutron powder diffraction experiment by Athauda et al.[31] confirmed this predicted *local* distortion within a single BiS$_2$ layer by observing indeed two different Bi-S bonds (with lengths of 2.68 Å and 3.11 Å). We note that the powder diffraction also suggested another antiferroelectric-like distortion[31], but such a mode is predicted to induce an instability at *M* point that is present in F-doped LaOBiS$_2$ but absent in the parent compound, and is thus excluded by single crystal diffraction very recently[32].

However, the in-plane *local distortions* predicted[11] and observed[31] do not define as yet a 3D structure in LaOBiS$_2$: there are two BiS$_2$ layers in a unit cell, and one needs to construct a 3D model of how the various 2D layers are stacked so as to minimize the energy. Yildirim[11] suggested a 3D structure that consists of two BiS$_2$ layers having the *same direction of distortion* (illustrated by the blue arrow shown in Fig. 1c) as a replacement of the unstable $T_0$ structure. This ferroelectric-like structure removes the phonon instability and thus lowers the energy, turning the 3D structure into non-centrosymmetric $T_1$ (space group *P2$_1$mn*) from centrosymmetric $T_0$ (*P4/nmm*)[11]. Interestingly, there are additional 3D stacking sequences of the 2D layers possible that were not previously examined in pristine LaOBiS$_2$ (without doping) by aligning the in-plane distortion in different directions. There are three possible stacking types that fulfill the local Bi-S1 distortion[11, 31]. In $T_1$ type, the local distortions of two BiS$_2$ layers are *along x* direction, in $T_2$ they occur along *x* and –*x* direction, while in $T_3$ are *x* and *y* direction. These distortions lower the system's symmetry from *P4/nmm* ($T_0$) to *P2$_1$mn* ($T_1$), or *P2$_1$m* ($T_2$) or *C2* ($T_3$). Among $T_1$, $T_2$, and $T_3$, only $T_2$ is centrosymmetric, whereas $T_1$ and $T_3$ are non-centrosymmetric. We next inquire as to the energetic order of stability of the different polytypes in LaOBiS$_2$ as well as in other compounds from the same general family.

### *III. Methods of calculation*

The geometrical and electronic structures are calculated by the projector-augmented wave (PAW) pseudopotential[33] and the generalized gradient approximation of Perdew,

Burke and Ernzerhof (PBE)[34] to the exchange-correlation functional as implemented in the Vienna *ab initio* package (VASP)[35]. The plane wave energy cutoff is set to 550 eV. Electronic energy minimization was performed with a tolerance of $10^{-6}$ eV, and all atomic positions were relaxed with a tolerance of $10^{-3}$ eV/Å.

We note that PBE functional does not take the long-range van der Waals interaction into account, and thus usually overestimate the interlayer space. To overcome this, we also use PBE-D2 exchange-correlation functional, with the implementation of van der Waals (VDW) correction, to optimize the distorted structures (see Sec. VI).

## *IV. Three levels of approximation for predicting bond relaxation and symmetry breaking*

The three polytypes differ in the stacking direction of electronic-active $BiS_2$ layers along the [001] direction. To gain insight into the various degrees of freedom we perform structural optimization in steps starting from the simplest and most restricted:

**Approach (a)**: If one allows the $T_0$ structure to relax but without imposing initially any symmetry breaking, only the *a, c* lattice vectors and four independent cell-internal atomic coordinate parameters (atomic positions of La, Bi, S1 and S2 sites along the *c* axis) can relax. This restricted relaxation is an artifact of the method of calculation, namely of assuming initially a high symmetry structure. It reflects the fact that the other degrees of freedom in $T_0$ that could relax are not permitted to do so because a high symmetry (zero force on atoms) is artificially imposed. As a result, the final structure in this relaxation method still has the same space group *P4/nmm* as the initial starting point, denoted by the $T_0$ point in Fig. 2. Although this is the only optimization treatment in most structure determination step in the DFT community, the result is incorrect because of the artificial restriction to high symmetry.

**Approach (b):** Next, we allow in-plane relaxation by initially lowering the symmetry but freeze all other degrees of freedom, while seeking the energy minimum under this constrain. To do so we start form $T_0$ but shift S1 atom of each $BiS_2$ layer along x or y direction, (depending on the $T_1$-$T_3$ type) and artificially *keep all other structural degrees of freedom frozen.* Such frozen Bi-S1 "bond length mismatch" distortion has been used by Ref. [11] for $T_1$ and is defined by the S1 displacement Δ. This is not the most general procedure, and is used here to observe what its physical consequence is. This simplified procedure of allowing just one mode to drive distortion is common in simple models of phase transitions. The results of this simplified model are the three curves in Fig. 2. They show that if one scans the energy as a function of $\Delta$, each of these structures has different energy minima shown as solid dots in $T_1$-$T_3$ curves. *This is clear evidence that in LaOBiS$_2$ $T_0$ structure is a saddle point and thus not able to exist stably.*

**Approach (c) – Most general**: The in-plane Bi-S1 distortion can enable additional degrees of freedom, including the lattice parameters and atomic positions. For example, $T_1$-type distortion will elongate the axis, while $T_2$-type distortion will twist the rectangular x-z plane of the unit cell into a parallelogram. Therefore, in the most general approach (c) we further relax all degrees of freedom starting from the energy-minimum structures of $T_1$-$T_3$, and get three final structures that are several meV/unit cell (u.c.) lower in energy, $T_1^*$, $T_2^*$ and $T_3^*$ indicated in Fig. 2 by diamond shapes. Unlike the $T_0$ structure, the $T_1^*$, $T_2^*$ and $T_3^*$ structures are thermodynamically stable. We also considered full relaxation starting from a generic symmetry breaking of the $T_0$ structure by randomly adding small distortions on both lattice vectors and atomic positions. We find from a few such starting points that the relaxed structures all become $T_1^*$, $T_2^*$ or $T_3^*$ structure, indicating that $T_1^*$-$T_3^*$ structure are the only three local minima around $T_0$ phase.

**Comparision between different fully relaxed polytypes:** Table I shows the total energy, space group, lattice parameter and Bi-S bond length of $T_0$, $T_1^*$-$T_3^*$ structure as well as the experimental refinement using $T_0$ structure. Several observations are found: (i) Although $T_0$ structure has a good agreement and a reasonable fit[6] (R factor < 0.1) with experiment regarding the in-plane lattice constant, there is only one kind of Bi-S bond, which is inconsistent with the results by neutron diffraction[31]. On the other hand, all of $T_1^*$ to $T_3^*$ suggests two kinds of in-plane Bi-S1 bonds, with the bond lengths 2.76-2.77 and 3.03-3.04 Å, respectively. These bond lengths agree with the experiments well[31]. (ii) With the presence of the in-plane distortion for $T_1^*$-$T_3^*$ structures relative to $T_0$, the buckling nature of the Bi-S1 plane is nearly unchanged. The calculated S1-Bi-S1 angle varies from 165° for $T_0$ to 164° for all $T_1^*$-$T_3^*$ structures, including one type of S1-Bi-S1 angle in $T_1^*$ and $T_2^*$, and two types in $T_3^*$. (iii) The $T_2^*$ structure is centrosymmetric and has the lowest energy, but only 1-2 meV/u.c. lower than the non-centrosymmetric $T_1^*$ and $T_3^*$. Such a small energy difference originates from the weak van der Waals interaction between adjacent $BiS_2$ layers. The total energies of the polytypes are so close that theoretical stability could be significantly affected by the temperature or crystal growing conditions. On the other hand, $T_0$ structure has the highest energy, about 10 meV/u.c. higher than $T_2^*$.

In order to visualize the transformation path from the saddle point $T_0$ to the stable structures (e.g., $T_1^*$ or $T_2^*$) in $LaOBiS_2$, we show in Fig. 3 the feasibility of converting $LaOBiS_2$ from $T_1^*$ to $T_2^*$ structure with $T_0$ structure as an intermediate state. In $T_0$ -> $T_1^*$ and $T_0$ -> $T_2^*$ path we start from $T_0$ structure and impose a tiny $T_1$- or $T_2$-type distortion on $T_0$. Then we record each step of DFT relaxation process from $T_0$ to $T_1^*$ ($T_2^*$) as the reaction coordinate, i.e., the geometric parameter that changes during the conversion between $T_1^*$ and $T_2^*$, by gradually changing the cell and the atomic position using conjugate gradient method. We find that $T_0$ structure stays at a saddle point, so it is unstable and will spontaneously relax to a distorted structure $T_1^*$ or $T_2^*$ (also probably $T_3^*$). On the other hand, the potential barrier between the transition of $T_1^*$ and $T_2^*$ structures is about 10 meV, which is not large enough to make $T_2^*$ as a unique structure. Instead, all $T_1^*$ to $T_3^*$ structures or their mixed phases are possible to happen in room temperature.

*V. Pressure-induced phase transition and polytypes in other $LaOBiS_2$-type compounds*

In order to unveil the physical origin of the polytypism, we investigated the ground-state structure of $LaOBiS_2$ under hydrostatic pressure. Figure 4(a) shows the in-plane distortion of $T_1$-$T_3$ structures presented by the Bi-S1 bond difference as a function of hydrostatic pressure. Zero pressure $P = 0$ corresponds to $T_1^*$-$T_3^*$ structures with PBE-calculated equilibrium lattice constant. We find that the pressure indeed suppresses the in-plane distortion for all $T_1$-$T_3$ structures. Above $P = 2.5$ GPa (corresponding an in-plane lattice constant $a = b = 4.02$ Å) the high-symmetry $T_0$ structure becomes the ground state of $LaOBiS_2$.

Following $LaOBiS_2$, many compounds having similar structure are reported for the purpose of exploring the $BiS_2$-based superconductivity[7, 10]. Therefore, it is interesting to ask if the other compounds in this family also have such polytypes or not, and what determines $T_0$ structure or distorted structures as a ground state. We substitute La (with Y), Bi (with Sb, As) or S (with Se, Te) site at one time and fix the other 3 sites the same as LaOBiS$_2$. Figure 4(b) shows the relative energy of the polytypes of these materials. We found that when substituting La with lighter elements or substituting S with heavier elements, the undistorted $T_0$ structures are the ground states, which means there is no polytypes in these materials. On the other hand, when substituting Bi with lighter elements Sb and As, $T_0$ structures are no longer dynamic stable (dash lines), while distorted $T_1^*$-$T_3^*$ are local minima. Among them $T_2^*$ is the ground state, with the total energy 10-15 meV lower than that of $T_1^*$. The energy differences between $T_2^*$ and $T_0$ for LaOSbS$_2$ and LaOAsS$_2$ are 167 and 718 meV, respectively, which is much larger than the counterpart of LaOBiS$_2$. Moreover, the in-plane M-S1 bonds (M: Bi, Sb or As) also reveals that LaOSbS$_2$ and LaOAsS$_2$ suffer larger distortion. The two As-S1 bond lengths in LaOAsS$_2$ are 2.43 Å and 3.31 Å. As a result, such large distortion breaks the 2D As-S square and forms individual zigzag As-S chains, as shown in the inset of Fig. 4(b).

From the study of pressure and various LaOBiS$_2$-type compounds we conclude that the polytypism in LaOBiS$_2$-family is mainly attributed to two reasons: the "substrate" effect of the intermediate layer (e.g., LaO) and hybridization effect of the active layer (e.g., BiS$_2$). In LaOMS$_2$ compounds (M: Bi, Sb, As), the mismatch between the optimum lattice constants of LaO (larger) and MS$_2$ (smaller) layers tends to stretch the M-S square and thus cause distortion. When the lattice of MS$_2$ layers becomes smaller without changing LaO layer, the stretching effect becomes stronger. Interestingly, although this effect originates from the interaction between LaO and MS$_2$ layers, the distortion in LaO layer is insignificant. It is because the bond energy of La-O is large so that the LaO network remains stable when suffering distortions. On the other hand, enlarging the lattice parameter of the side layer or reducing the lattice parameter of the intermediate layer induce better lattice match between side and intermediate layer so that the distortion is unlikely to happen, as the case in LaOBiSe$_2$, LaOBiTe$_2$ and YOBiS$_2$. For LaOBiS$_2$ under pressure, the BiS$_2$ and LaO layers are simultaneously compressed, in which case the enhanced Bi-S1 bond strength could stabilize the $T_0$ structure and prevent distortion.

The other factor that can lead to distortion is within the local structure of the side layer. For the trend from Bi to As, the group-V cations behave more covalent character while less ionic character. As a result, they are inclined to form $sp^3$ hybridization with 3 S atoms and the other bond is the lone-pair electron. The lone-pair bond usually has larger repulsive force than the other bonds, and thus breaks two M-S1 bonds to from zigzag chains.

## *VI. The results of VDW correction*

VDW interactions represent the interaction of an **induced moment** (dipole, quadropole, etc) with itself. This is strictly zero in mean field Hartree -Fock or DFT. From Table I we find that DFT without VDW gives the length of c axis ($T_0$ and $T_1^*$-$T_3^*$) as 14.25 Å, much larger than the observed 13.83 Å. This is because the PBE functional didn't take the long-range van der Waals interaction into account, and thus usually overestimate the interlayer space.

To overcome this, we also use PBE-D2 exchange-correlation functional that includes a rough VDW term to optimize the distorted structures. We find that the *a* and *c* axis of $T_0$ structure are 3.99 Å and 13.62 Å, respectively. Although PBE-D2 gives a better agreement with experiments on c axis (13.83 Å), it underestimate the *a* axis by 1.5% than that of experiments or PBE results (4.05 Å). More importantly, because of the biaxial strain effect within *x-y* plane under PBE-D2 method, the undistorted structure $T_0$ becomes the ground state. We performed the relaxation starting from the PBE relaxed $T_1^*$-$T_3^*$ structure and find that the final configuration is the $T_0$ structure. Phonon calculations of LaOBiS$_2$ for smaller in-plane lattice parameters also suggested that the undistorted structure $T_0$ is dynamic stable[11]. Since the accuracy of *a* axis is more important to investigate the in-plane distortion, we used PBE functional for the base of calculation and analysis.

## *VII. Summary*

By using first-principles total energy and stability calculations, we predict that the local Bi-S distortion leads to three polytypes of LaOBiS$_2$-type materials according to the different stacking pattern of BiS$_2$-like layers. The in-plane distortion, mainly induced by the interaction between the intermediate and the side layers and the electron hybridization in the side layer, can lower the total energy comparing with the conventional accepted *P4/nmm* structure. In LaOBiS$_2$, the energy difference between the polytypes is small, indicating the possible coexistence of all polytypes in the real sample. While we cannot determine if the superposition is dynamic in nature (time-dependent doping, we note that such a superstructure with the combination of $T_1$, $T_2$ and $T_3$ polytypes will effectively lower the symmetry to the P1 space group. Such a superstructure symmetry has been recently suggested by a single-crystal neutron-diffraction experiment of Athauda *et al*.[32] Finally, external pressure and different choices of the intermediate atoms (replacing La) or active atoms (BiS$_2$) manifest different ground state polytypes. One can thus tune the distortion and the ground state by both physical and chemical means. Our findings provide a clear picture on the complexity of the crystal structure of LaOBiS$_2$-type materials, which might be relevant to the mechanism of the recently found BiS$_2$-based superconductivity.

**Acknowledgements**

We are grateful for the helpful discussions with Prof. Daniel Dessau from University of Colorado Boulder and Prof. Despina Louca from University of Virginia. This work was supported by NSF Grant titled "Theory-Guided Experimental Search of Designed Topological Insulators and Band-Inverted Insulators" (No. DMREF-13-34170). This work used the Extreme Science and Engineering Discovery Environment (XSEDE), which is supported by National Science Foundation grant number ACI-1053575.

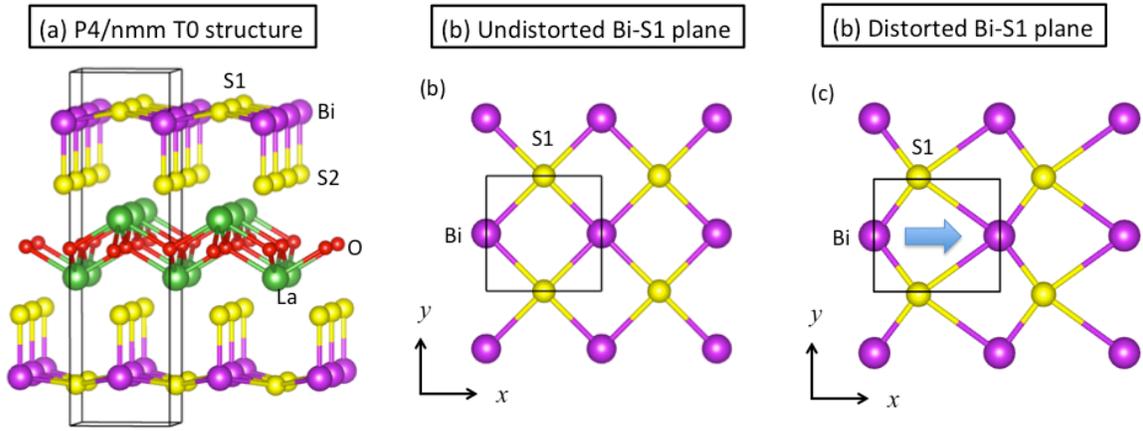

Fig. 1: (a) P4/nmm $T_0$ structure of LaOBiS$_2$. (b) Top view of Bi-S1 2D square consisting of Bi and S1 atoms in $T_0$ structure. (c) Top view of Bi-S1 2D square with distortion. The blue arrow along $x$ direction indicates the displacement of S1 and the spontaneous polarization. The black frame indicates the unit cell.

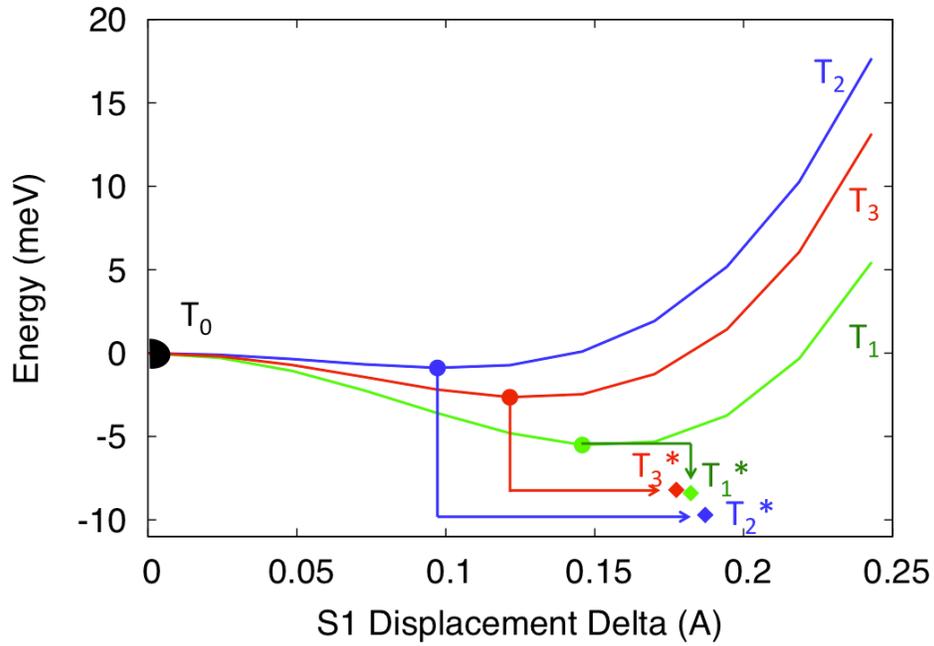

Fig. 2: Total energy as a function of S1 displacement in $T_1$-$T_3$ stacking types of LaOBiS$_2$. $T_1^*$-$T_3^*$ indicate the complete relaxed structures of $T_1$-$T_3$ type (approach c), respectively.

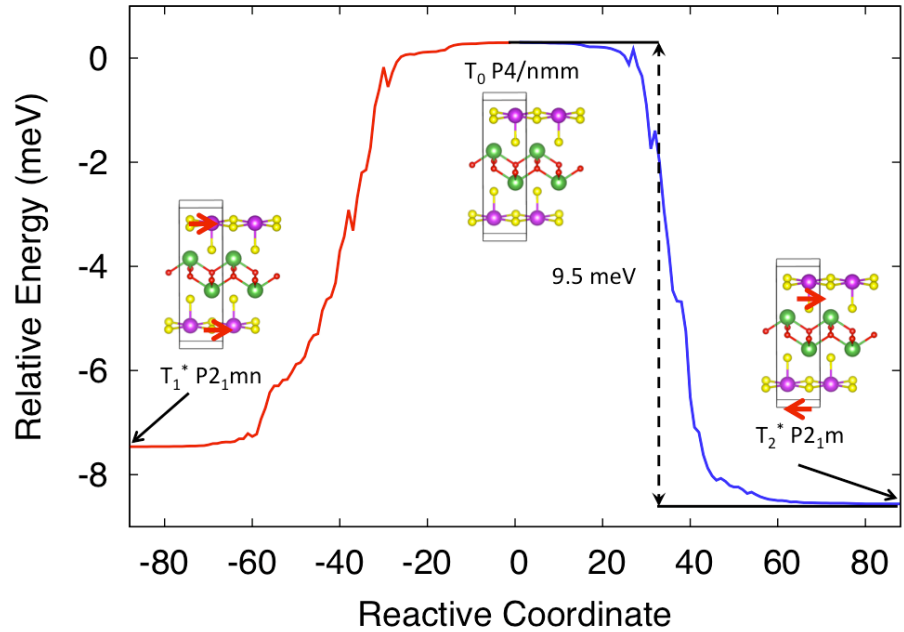

Fig. 3: Total energy change during the transformation from $T_1^*$ to $T_2^*$ structure with $T_0$ structure as an intermediate state. The reaction coordinate in *x*-axis is presented by the DFT relaxation steps from $T_0$ to $T_1^*$ and $T_2^*$. The red and blue lines denote the transition from $T_0$ to $T_1^*$ and $T_2^*$, respectively. The red arrows in the crystal structure denote the direction of the $BiS_2$ in-plane distortion.

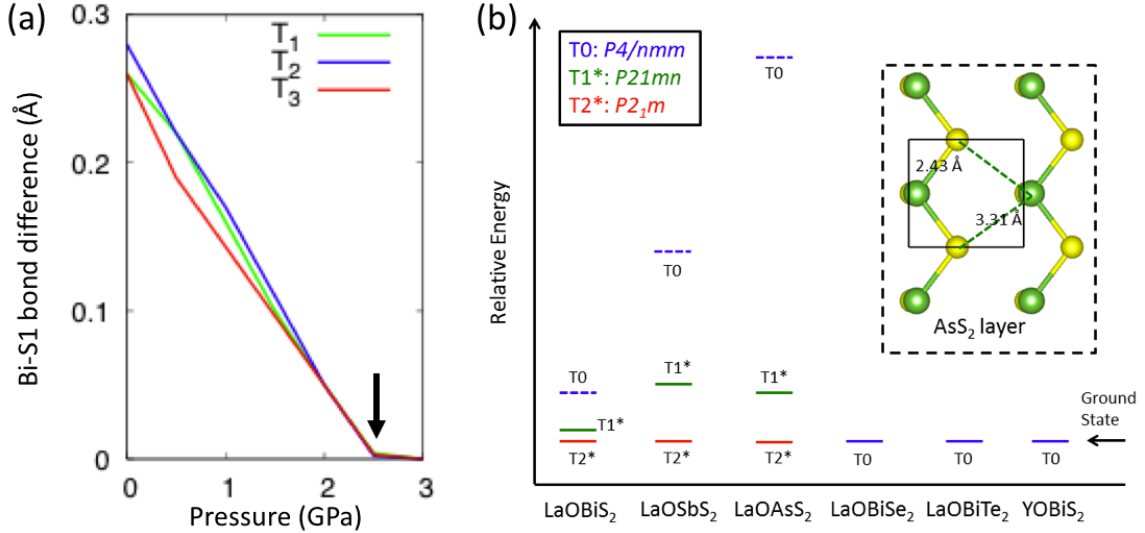

Fig. 4: (a) The difference of bond length within Bi-S1 plane of distorted $T_1$-$T_3$ structures as a function of hydrostatic pressure. The black arrow corresponds to 2.5 GPa where the Bi-S1 bond length become isotropic, indicating a phase transition from a distorted structure to high-symmetry $T_0$ structure. Since the difference between a distorted $T_1$-$T_3$ and $T_0$ structure is just a stretched Bi-S1 plane (but no bond breaking), different pressure will lead to a structure with either larger or smaller in-plane distortion. If the distortion vanishes under pressure, then a phase transition will ensue from $T_1$-$T_3$ to $T_0$. (b) Schematic illustration of the relative energy of different polytypes in LaOBiS$_2$-family. The dashed horizontal lines denote structures that are dynamic unstable. Dashed box: top view of AsS$_2$ layer in LaOAsS$_2$, in which the unit cell is denoted by the solid frame. The strong distortion separates the layer into individual zigzag chains.

Table I: The stacking configurations of in-plane distortion, space group, inversion symmetry, total energy per u.c., Bi-S bond lengths and lattice parameters of LaOBiS$_2$, including experimental results and DFT calculations $T_0$ and $T_1^*$-$T_3^*$ on PBE level. The total energy of $T_0$ structure is set to zero as a reference.

| | Expt.[31] | $T_0$ | $T_1^*$ | $T_2^*$ | $T_3^*$ |
|---|---|---|---|---|---|
| Stacking of distortion | --- | No distortion | (x, x) | (x, -x) | (x, y) |
| Space group | *P4/nmm* | *P4/nmm* | *P21mn* | *P2$_1$m* | *C2* |
| Inversion symmetry | Unknown | Yes | No | Yes | No |
| Energy (meV) | --- | 0 | -8.4 | -9.7 | -8.2 |
| Bi-S bond (Å) | 2.68/ 3.11 | 2.89/ 2.89 | 2.77/ 3.03 | 2.76/ 3.04 | 2.77/ 3.03 |
| a (Å) | 4.054 | 4.050 | 4.071 | 4.073 | 4.060 |
| b (Å) | 4.054 | 4.050 | 4.051 | 4.051 | 4.060 |
| c (Å) | 13.825 | 14.250 | 14.266 | 14.273 | 14.266 |